\documentclass[reprint,nofootinbib,amsmath,amssymb,aps]{revtex4-2}

\usepackage{graphicx}% Include figure files
\usepackage{dcolumn}% Align table columns on decimal point
\usepackage{bm}% bold math
\usepackage{hyperref}% add hypertext capabilities

\begin{document}

\title{Fractons, symmetric gauge fields and geometry}

\author{Francisco Pe\~na-Ben\'itez}
 \affiliation{Max Planck Institute for the Physics of Complex Systems and W\"{u}rzburg-Dresden Cluster of Excellence ct.qmat. N\"othnitzer Str. 38, 01187 Dresden, Germany.\\
 Institute of Theoretical Physics, Wroc\l aw University of Science and Technology, 50-370 Wroc\l aw, Poland.}

%%%%%%%%%%%%%%%%%%%%%%%%%%%%%%%%%%%%%%%%%%%%%%%%%%%%%%%%%%%%%%%%%%%%%%%%%%%%%%%%%%

\begin{abstract}
Gapless fracton phases are characterized by the conservation of certain charges and their higher moments. These charges generically couple to higher rank gauge fields. In this paper we study systems conserving charge and dipole moment, and construct the corresponding gauge fields propagating in arbitrary curved backgrounds. The relation between the symmetries of these class of systems and spacetime transformations is discussed. In fact, we argue that higher rank symmetric gauge theories are closer to gravitational fields than to a standard gauge theory.
\end{abstract}

%%%%%%%%%%%%%%%%%%%%%%%%%%%%%%%%%%%%%%%%%%%%%%%%%%%%%%%%%%%%%%%%%%%%%%%%%%%%%%%%%%

\maketitle

In modern physics, symmetries are a fundamental paradigm to organise degrees of freedom of a given system. Typically, the symmetries can be divided in terms of internal, and spacetime symmetries. For instance, electric charge and isospin conservation are consequence of the former, whereas conservation of momentum and angular momentum of the latter. Understanding the distinction between the two classes of transformations is vital in order two characterize the force fields associated to interacting charged matter. At the fundamental level, matter charged under internal symmetries interact via gauge fields, whereas gravitational fields carry the "force" between fields charged with respect to spacetime symmetries, e.g. energy (mass), and/or momentum. Although physicist have made several attempts to describe gravity as a gauge theory, and a successful algorithm to gauge (relativistic or not) spacetime symmetries has been developed \cite{ivanov1982gauge,grignani1992gravity,son2006general,andringa2011newtonian}, strictly speaking gravitational fields are not gauge fields.

On the other hand, in recent years, a new class of matter excitations has been proposed named as fractons \cite{Fractons1,Pretko:2020cko}. The main feature a quasiparticle needs to show to be called fractonic is the property of reduced mobility\footnote{Do not confuse with the fractons introduced in  \cite{Khlopov:1981wm} in the context of nuclear interactions.}. In fact, fracton matter can be classified in terms of gapless \cite{Xu2006,Xu2010,Pretko2017spinliquid,Pretko:2016lgv,You_emergent_2020,pretko2019crystal} and gapped \cite{Chamon:2005,Haah2011,Vijay:2016phm,Bravyi:2011,Williamson2016} phases. In particular, gapless fracton phases are described by the conservation of certain charges and their higher moments. This peculiar behavior has as consequence a non-standard continuity equation (see Eq. \eqref{eq_scalconserv} for the case of charge and dipole conservation, and \cite{Pretko:2016lgv,Gromov_multipole_2019} for more general examples). This class of symmetries naturally arise in the context of spin liquids \cite{Pretko:2016lgv,You_emergent_2020,pretko2019crystal}, quantum Hall and elasticity \cite{Pretko:2018fed,Gromov2019elastic,Nguyen:2020yve,du2021volume}, topological defects \cite{Nguyen:2020yve,Aasen:2020zru,Doshi2021}, and beyond the condensed matter realm in systems with Galileons \cite{Nicolis:2008in}. 

For simplicity,  we shall focus on the case of systems preserving a scalar charge and its corresponding dipole moment. This symmetry has the form of some generalised $U(1)$ symmetry. Actually, is not hard to conclude that the interaction among such type of charges should be carried by generalized 'electromagnetic gauge fields' with the spatial vector potential substituted by a symmetric tensor field \cite{Pretko:2016lgv,Pretko2017spinliquid}. However, if we try to follow the standard minimal coupling rule to couple charged matter to these gauge fields we run into problems, and the only way out proposed so far  is with a non-Gaussian theory \cite{Gromov_multipole_2019,Yuan:2019geh}, making analytic computations really hard. Nonetheless, certain progress has been made in the  hydrodynamic description for such systems in absence of gauge fields \cite{Yuan:2019geh,GromovLucas2020,grosvenor2021hydrodynamics,glorioso2021breakdown,argurio2021fractons}. 

Furthermore notice, that similarly to angular momentum, the value of a dipole moment depends on the location of the origin of the coordinates system, making it hard to link its conservation to an internal symmetry group, indicating that charge and dipole conservation, should be related to a spacetime symmetry group, rather than to a purely internal one. This observation was one of the main motivations for the study presented here. Also notice that the fractonic 'force' field is a symmetric tensor in similarity with the metric which is the responsible for gravitational forces. Actually, in \cite{pretko2017emergent,Gromov2019elastic} a connection between fractons dynamics and linearized gravity has been discussed. In this paper, we propose a geometric theory where the symmetric gauge fields play the role of vielbeins in a vertical space to the physical spacetime. This construction pave the road to a systematic understanding of more generic multipole preserving gauge theories, and opens up a path for the construction of low energy, and possibly microscopic, fractonic matter actions.

Our first result relies on the interpretation of the Nambu-Goldstone mode in a spontaneously broken phase, as the embedding coordinate of the physical spacetime in a higher dimensional Heisenberg spacetime \cite{calin2005geometric,vukmirovic2015classification}. In fact, the fracton transformations belong to the isometry group of such spacetime. Given this interpretation, a minimal (non-linear) theory for the Nambu-Goldstone field has the form of a Born-Infeld action [see Eq. \eqref{eq_SSB}].
The second and main result Eq. \eqref{eq_gaugeAction} is a fully diffeomorphism and gauge invariant higher rank  gauge theory. Actually, the construction suggests that symmetric gauge fields will generically become massive on a curved background, and to preserve the gauge invariance we need to introduce a Stueckelberg field. In particular, our theory reduces to the models proposed in \cite{Pretko:2016lgv} once the spacetime is assumed to be flat. 

The paper is organized as follows: In Sec. \ref{sec:charge-dipole_conservation} we describe the conservation laws and symmetry algebra of a system conserving energy, mometum, angular momentum, and a scalar charge with its corresponding dipole moment. Next, in Sec. \ref{sec:Spontaneous_Symmetry_Breaking} we discuss the relation of the symmetries of the system with the so-called Heisenberg group, and study the spontaneous symmetry breaking of fracton charges. In Sec. \ref{sec:Gauge_Theory}, we gauge the full symmetry group obtaining a generalization of the fracton electrodynamic gauge theories on curved spacetimes. Then, in Sec. \ref{sec:Discussion} we discuss the outputs of our proposal, possible implications, and outlooks.

%%%%%%%%%%%%%%%%%%%%%%%%%%%%%%%%%%%%%%%%%%%%%%%%%%%%%%%%%%%%%%%%%%%%%%%%%%%%%%%%%%
%%%%%%%%%%%%%%%%%%%%%%%%%%%%%%%%%%%%%%%%%%%%%%%%%%%%%%%%%%%%%%%%%%%%%%%%%%%%%%%%%%
\section{Charge-Dipole conservation}\label{sec:charge-dipole_conservation}
%%%%%%%%%%%%%%%%%%%%%%%%%%%%%%%%%%%%%%%%%%%%%%%%%%%%%%%%%%%%%%%%%%%%%%%%%%%%%%%%%%
%%%%%%%%%%%%%%%%%%%%%%%%%%%%%%%%%%%%%%%%%%%%%%%%%%%%%%%%%%%%%%%%%%%%%%%%%%%%%%%%%%

Gapless fracton phases are characterized by the conservation of certain charges and their higher moments. The simplest case, corresponds with the conservation of a charge $Q$ and its dipole $Q^a$,  which in $n$ space dimensions, at the macroscopic level, can be formulated in terms of the charge density $\rho$ as  
\begin{equation}
\frac{ d}{ dt}\int d^n x \left(\lambda  + \boldsymbol{\beta}\cdot \mathbf x\right)\,\rho =0 \,,\label{eq_monopole}
\end{equation}
with $\lambda,\boldsymbol{\beta}$ arbitrary parameters.
In a system with such conservation law, charges are immobile, whereas dipoles can freely move.
In fact, similarly to what happens with momentum and angular momentum\footnote{In a system with both momentum and angular momentum conservation once momentum is conserved, conservation of angular momentum follows.}, both charges are conserved once the single (generalized) continuity equation 
\begin{equation}\label{eq_scalconserv}
\partial_t\rho+\partial_a\partial_bJ^{ab}=0\,,\qquad a,b=1,2,\ldots,n\,,
\end{equation}
is satisfied. The distinguishing feature in this class of systems is that charge is relaxed via a tensorial current. An immediate consequence of such conservation law, is that a gauged version of the symmetry would require the presence of gauge fields $A_0,A_{ab}$ with the transformation rule $A_0\to A_0-\partial_t\lambda$, and $A_{ab}\to A_{ab}+\partial_a\partial_b\lambda$, and the 'gauge fields' coupling to the fractonic matter as follows
\begin{equation}
S = S_0[A_0,A_{ab}]+\int\,d^{n+1}x\left(\rho A_0 + J^{ab}A_{ab} \right)\,.
\end{equation}
Such type of theories have been proposed as a generalization to electrodynamics \cite{Pretko:2016lgv}. However, due to the unusual transformation law of the fields, it is not clear in what sense they are actual gauge theories. In addition, from this perspective, it is not obvious whether it is possible to put the theory on a curved manifold without spoiling the gauge symmetry \cite{slagle2019symmetric}.

In order to understand the tension between the spacetime transformations and the gauge symmetry introduced above it is useful to notice that the dipole charge $Q^a$ is charged under spatial translations. The main reason, is that its value will change once the origin of the space is shifted, contrary to what happens to the charge $Q$, which is insensitive to the location of the origin. This is an unusual property for internal symmetries. In fact, a careful analysis of the action of time and space translations, rotations, and the transformations generated by the fracton charges with generators $H,P_a,L_{ab},Q^a,Q$ respectively, imply that the whole set of transformations form a continuous Lie group $\mathcal G$ with its corresponding Lie algebra satisfying the non-vanishing Lie brackets  \cite{grosvenor2021hydrodynamics}
\begin{align}\label{eq_algebra}
[P_a,Q^b] &= \delta_a^b Q\,,\\\nonumber
[P_a,L_{bc}] &= \delta_{ac} P_b - \delta_{ab} P_c  \,,\\\nonumber
[Q^a,L_{bc}] &= \delta^a_c Q_b - \delta^a_b Q_c  \,, \\\nonumber
[L_{ab}, L_{bc} ] &= \delta_{ac}L_{bd} - \delta_{ad}L_{bc} - \delta_{bc} L_{ad} + \delta_{bd}L_{ac}\,.\nonumber
\end{align}
This algebra makes evident that conservation of charge and dipole are consequence of a spacetime symmetry group. Contrary to the usual case of $U(1)$ charges. A similar example is the case of mass conservation in Galilean invariant theories \cite{andringa2011newtonian}.  Actually, Eqs. \eqref{eq_algebra} show similarities with the Bargmann algebra once the generator of Galilean boosts is identified with the dipole generator $Q^a$, and mass with charge $Q$. See also \cite{hartong2015gauging,copetti2020torsion} for the similarities with Carroll theories.

%%%%%%%%%%%%%%%%%%%%%%%%%%%%%%%%%%%%%%%%%%%%%%%%%%%%%%%%%%%%%%%%%%%%%%%%%%%%%%%%%%
%%%%%%%%%%%%%%%%%%%%%%%%%%%%%%%%%%%%%%%%%%%%%%%%%%%%%%%%%%%%%%%%%%%%%%%%%%%%%%%%%%
\section{Group manifold and spontaneous symmetry breaking}\label{sec:Spontaneous_Symmetry_Breaking}
%%%%%%%%%%%%%%%%%%%%%%%%%%%%%%%%%%%%%%%%%%%%%%%%%%%%%%%%%%%%%%%%%%%%%%%%%%%%%%%%%%
%%%%%%%%%%%%%%%%%%%%%%%%%%%%%%%%%%%%%%%%%%%%%%%%%%%%%%%%%%%%%%%%%%%%%%%%%%%%%%%%%%
 
  In this section we shall give a geometric interpretation to the group $\mathcal G$ in terms of Heisenberg spaces \cite{calin2005geometric,vukmirovic2015classification}. In particular, the Heisenberg algebra has $2n+1$ dimensions, and non-vanishing brackets
	\begin{equation}
		\label{eq_Heialg} [P_a,Q^b] = \delta_a^b Q.
	\end{equation}
	Therefore, it is a subalgebra of the entire fractonic algebra introduced above (see Eq. \eqref{eq_algebra}). In order to get some intuition on the relation of the Heisenberg group with an actual spacetime, we recall that the $n-$dimensional real space with additive composition  is the coset space of the Euclidean and rotations groups, i.e. $\mathbb R^n=E_n/SO(n)$. In full analogy, we define the (fractonic) Heisenberg space $\mathbb H_{2n+1,1}=\mathcal G/SO(n)$.
	
	To construct such space, we parametrize  elements of the coset with coordinates $(y^0,y^a,z_a,\phi)$ via the exponential map 
	\begin{equation}\label{eq_Omega}
		\Omega=e^{y^0H+y^aP_a}e^{z_aQ^a}e^{\phi Q},
	\end{equation}
where $y^0$, and $y^a$ are internal coordinates showing certain resemblance with the comoving  time and "fluid"  elements respectively used in fluid dynamics  \cite{Dubovsky_2012}. On the other hand, $z_a$, and $\phi$ are the Nambu-Goldstone fields parametrizing the spontaneous breaking of the generators $Q,Q^a$. 

Using the left action of the group on itself, we define the transformed element of the coset as $\tilde \Omega=g\Omega e^{-\beta^{ab}L_{ab}}$, with $g=e^{\zeta^0H+\zeta^aP_a}e^{\beta_aQ^a+\lambda Q}e^{\beta^{ab}L_{ab}}$. The infinitesimal transformations read
	\begin{align}\label{eq:fract_trans}
		\delta y^{0}&= \zeta^0\,,\qquad \delta y^{a}= y^b\beta_b\,^a + \zeta^a\,,\\
	\nonumber	\delta z_a &= z_b\beta^b\,_a + \beta_a\,\quad \delta \phi=\lambda-\beta_ay^a\,.
	\end{align}
The Maurer-Cartan form 
$\mathcal A=\Omega^{-1}d\Omega$ reads
\begin{equation}
\mathcal A	 = \tau H + e^a P_a + \omega_a Q^a + v Q,
\end{equation}
with $\tau=dy^0,e^a=dy^a,\omega_a=dz_a, v=d\phi + z_ady^a$. The Maurer-Cartan equations imply $dv = \omega_a\wedge e^a$. In addition, the (invariant) inverse vector fields are 
 \begin{equation}\label{eq_invariant-basis}
	t = \frac{\partial}{\partial y^0}\,,\quad     \mathcal E_a = \frac{\partial}{\partial y^a} - z_a\frac{\partial}{\partial \phi}\,,\quad
	\mathcal{\bar E}^a = \frac{\partial}{\partial z_a}\,,
	\quad  V  = \frac{\partial}{\partial \phi}\,.
\end{equation}
Notice, that they define a basis where $\mathcal E_a$, $\mathcal{\bar E}^b$ do not commute. In particular, their Lie bracket is $[\mathcal E_a,\mathcal{\bar E}^b] =  \delta^b_aV$, which corresponds with  the Heisenberg Lie algebra  Eq. \eqref{eq_Heialg}, if we identify $\mathcal E_a\to P_a$, $\mathcal {\bar E}^a\to Q^a$, and $V\to Q$. From now on, we will call this space the Heisenberg spacetime $\mathbb H_{2n+1,1}$. 

In addition, we introduce the physical spacetime $\mathbb M_{n+1}$ as the base space of a fibre bundle with total space being $\mathbb H_{2n+1,1}$. The projection from $\mathbb H_{2n+1,1}$ to $\mathbb M_{n+1}$ can be fixed as $y^0=x^0$, $y^a=\delta^a_\mu x^\mu$, this gauge fixing identifies the base manifold coordinate transformations with parameters $\xi^\mu$ to the internal translations and rotations such that
\begin{equation}
	\xi^\mu = \zeta^\mu + \delta^\mu_a\delta_\nu^b\beta^a\,_b x^\nu\,.
\end{equation}
In particular, the scalar field $\phi$ now transforms as
\begin{equation}
	\delta\phi = \lambda - \beta_a x^a + \zeta^\mu\partial_\mu\phi   - \frac{1}{2}\beta^{ab} (x_a\partial_b-x_b\partial_a)\phi\,,
\end{equation} 
also notice that this identification removes the distinction between internal and base spacetime indices. A last necessary gauge fixing corresponds with so-called inverse Higgs constraint \cite{isham1971nonlinear,low2002spontaneously,kharuk2018solving}. One of the indications we should introduce it in our system, is the fact that $-\partial_a\phi$ and $z_a$ have the same transformation property, suggesting that these fields are not independent. Therefore, we remove redundant modes by setting $z_a=-\partial_a\phi$. Finally, after the gauge fixing we obtain the covariant derivatives of the Nambu-Goldstone field are
\begin{equation}
	v_0 = \partial_0\phi\,,\quad
	\omega_{0 a} =-\partial_a\partial_0\phi\,,\quad
	\omega_{a b} = - \partial_a\partial_b\phi \, .
\end{equation}

Moreover, we can introduce on $\mathbb H_{2n+1,1}$ the metric $\bar G_{2n+2} = p\tau^2+ (e^a)^2+(\omega_a)^2+v^2$, with $p$ a sign that we will fix below. In fact, a possible interpretation, is that the spontaneously broken phase is captured by the localization of a $n+1-$dimensional "membrane" at the points $(x^\mu,-\partial_a\phi(x),\phi(x))$ . With this embedding, the induced volume reads $vol_{n+1}=\sqrt{|\bar G_{n+1}|}d^{n+1}x$. 

Having constructed the proper invariants of the system,  we can write the most general  low-energy effective action for the spontaneously broken phase as
\begin{equation}\label{eq_SSB1}
	S_{SSB}=\int d^{n+1}x\sqrt{|\bar G_{n+1}|}\,\mathcal L\left(v_0,(\omega_{0a})^2,(\omega_{ab})^2\right)\,.
\end{equation}
In general, the form of  the effective Lagrangian will depend on the precise microscopic system we consider. However, 
it is interesting to notice that the minimal theory with $\mathcal L = -\alpha$ have an action of the Dirac-Born-Infeld form
  \begin{equation}\label{eq_SSB}
	S_{SSB}=-\alpha\int d^{n+1}x\sqrt{|\mathrm{det}\,\left(p\delta_\mu^0\delta_\nu^0 + \delta_\mu^a\delta_\nu^a+ B_{\mu\nu}\right)|}.
\end{equation}
with
\begin{align}\label{eq_Bfield}
    B_{\mu\nu}(x) &= v_\mu(x) v_\nu(x) + \omega_{\mu a}(x)\omega_{\nu a}(x)\,.
\end{align}
 Actually, if we assume gradients are small, and introduce the derivative expansion $\partial_0 \sim \nabla^2$ the action reads
  \begin{equation}
      S_{SSB} \approx -\alpha\int \left(1 + \frac{p}{2}(\partial_0\phi)^2 + \frac{1}{2}(\partial_a\partial_b\phi)^2 + \ldots \right)  \,,
  \end{equation}
  notice that setting $p=-1$, will guarantee a positive definite energy for the linearized theory. In next section, we will see that such condition will also give the right signs in the action for the symmetric gauge fields.
  
  It is important to emphasize that, although Eq. \eqref{eq_SSB} has a nice geometric interpretation, because its equation of motion will extremize the volume of the orbits of the internal coordinates, in general the precise form of the action should be model dependent. From now on, we will call $G_{\mu\nu}=-\tau_\mu\tau_{\nu} + e_\mu\,^a e_\nu\,^a$ the spacetime metric and $B_{\mu\nu}$ the fracton metric. %In the next section we will generalize this construction to the case of curved spacetimes, and local fracton transformations.

%%%%%%%%%%%%%%%%%%%%%%%%%%%%%%%%%%%%%%%%%%%%%%%%%%%%%%%%%%%%%%%%%%%%%%%%%%%%%%%%%%
%%%%%%%%%%%%%%%%%%%%%%%%%%%%%%%%%%%%%%%%%%%%%%%%%%%%%%%%%%%%%%%%%%%%%%%%%%%%%%%%%%
\section{The gauge theory}\label{sec:Gauge_Theory}
%%%%%%%%%%%%%%%%%%%%%%%%%%%%%%%%%%%%%%%%%%%%%%%%%%%%%%%%%%%%%%%%%%%%%%%%%%%%%%%%%%
%%%%%%%%%%%%%%%%%%%%%%%%%%%%%%%%%%%%%%%%%%%%%%%%%%%%%%%%%%%%%%%%%%%%%%%%%%%%%%%%%%

Previous formulation naturally allows us to gauge the fractonic symmetry in full analogy with gravitational theories\footnote{See \cite{garcia2019gauged,copetti2020torsion} for similar constructions, although different physical systems.}. This procedure should provide a consistent field theory for fractonic gauge fields in curved backgrounds. In order to do so, we will follow the method described in \cite{ivanov1982gauge}. Using this technique the spacetime coordinates are interpreted as the Stueckelberg fields associated to breaking of local translations. Nonetheless, in our system we have embedded the spacetime in a larger space, therefore, in full analogy we could expect that local fracton translations might be broken with Stueckelberg fields ($z_a(x),\phi(x)$).
In such regime the connection must be defined as $\mathcal A = \Omega^{-1}(d + \tilde{\mathcal A})\Omega$, with $\Omega$ defined in Eq. \eqref{eq_Omega}, and $\tilde{\mathcal A}$ being the corresponding gauge field. Since the components of $ \mathcal{\tilde A}$ along the algebra directions are independent,  we find convenient to parametrize them as
\begin{equation}
   \tilde{ \mathcal A} =  \tilde\tau H + \tilde e^a P_a + \tilde \omega_a Q^a + (\tilde v + y^a\tilde\omega_a )Q + \frac{1}{2}\omega^{ab}L_{ab}\,. 
\end{equation}
By construction $\mathcal A$   will be invariant with respect to the broken generators. On the other hand, it will transform as a gauge field with respect to the unbroken generators, and reads
\begin{align}
\mathcal A &=  \tau H + e^a P_a + \omega_aQ^a + v Q + \frac{1}{2}\omega^{ab}L_{ab}\,,
\end{align}
with  $\tau=dy^0+\tilde \tau$, $e^a = Dy^a + \tilde e^a$, $\omega_a=Dz_a+\tilde\omega_a $,  $v=d\phi + z_ae^a + \tilde v$, and  the covariant exterior derivative $Dp^a=dp^a-\omega^{a}\,_b\wedge p^b$. In particular, we shall interpret $(\tau,e^a)$ as a local basis with inverse vielbeins $(t^\mu$, $E^\mu\,_a)$ respectively, and $\omega^{ab}$ as the spin connection of the spacetime. With them we define the spacetime metric  $G_{\mu\nu}= e_\mu\,^{a}e_\nu\,^{a} - \tau_\mu\tau_\nu $, and  the fracton metric $B_{\mu\nu}=v_\mu v_\nu + \omega_{\mu a}\omega_{\nu a}$.  Fields with internal space indices $a,b,c$ transform as vectors (tensors) with respect to local $SO(n)$ transformations, whereas the spin  connection as a non-Abelian gauge field (see Appendix \ref{App_GaugeTrans}).

 Under an infinitesimal internal translation  $g(x) = 1 + \beta_a(x) Q^a + (\lambda(x) + y^a\beta_a(x) ) Q$, the Stueckelberg  and gauge fields transform as
 \begin{align}
 	&\delta\phi=\lambda\,,\quad \delta z_a =\beta_a\,,\quad  \delta\omega^{ab}=0\,,\\
 	&\delta\tilde\omega_a = -D\beta_a \,,\quad   \delta \tilde v  = -d\lambda - e^a\beta_a\,.\label{eq_transomegav}
 \end{align}
As in the previous section we identify the internal and physical coordinates as $(y^0=x^0 , y^a=\delta^a_\mu x^\mu)$ and fix the Stueckelberg to be  $z_a = -E^\mu\,_a\partial_\mu\phi$, this condition requires $\beta_a=-E^\mu\,_a\partial_\mu\lambda$. After doing so, we notice that $E^\mu\,_a\tilde v_\mu$ will not transform under fracton gauge transformations. Therefore, we set them to zero ($\tilde v= A_0\tau$) because those components will not couple to the fractons current.

The curvature two-form of the theory is then defined as $\mathcal R=d\mathcal A+\mathcal A\wedge \mathcal A$ which can be expanded as
\begin{align}
\nonumber\mathcal R = &d\tau H + De^aP_a +  \frac{1}{2}R^{ab}L_{ab} + (F_a +  z_b R^b\,_a) Q^a +\\
&   (f + z_a De^a) Q\,,
\end{align}
where $R^{ab}=D\omega^{ab}$ is the curvature associated to the spin connection,  and the fractons field strengths
\begin{equation}
F_a = D\tilde\omega_a  \,,\qquad \label{eq_internalTorsion}f = d\tilde v +  e^a\wedge \tilde\omega_a \,.%= dA_0\wedge\tau + A_0d\tau \,. 
\end{equation}
 $F_a$ and $f$ are not invariant under internal translations if the curvature $R^{ab}$ and the spatial torsion $De^a$ are not vanishing (see Appendix \ref{App_GaugeTrans}), which justify the presence of the Stueckelberg fields. From this perspective, the breaking of dipole conservation has the same origin as the breaking of translational invariance in curved spacetime \cite{ivanov1982gauge,grignani1992gravity}. In odd spacetime dimensions local translations can be preserved, and gravitational theories can be related to Chern-Simons models with Poincar\'e as gauge group \cite{witten19882+}. Nonetheless, studying the generalization to our problem goes beyond the scope of this paper, and shall be left for future studies. 

For simplicity, we assume the spacetime to be torsionless ($d\tau=De^a=0$). Such constraints, fix $\tau$ to be a closed form, and allow us to express the spin connection in terms of the vielbeins. In the Appendix \ref{App_constraints} the spin connection in terms of the vielbeins is shown. When the torsion vanishes $f$ becomes gauge invariant, and for convenience  we parametrize the dipole gauge field as
\begin{equation}
	\tilde\omega_a = (\theta_a - E^\mu\,_a\partial_\mu A_0)\tau + C_{ba}e^b +  A_{ab}e^b\,,
\end{equation}
with $ A_{ab}= A_{ba}$, and $C_{ab}=-C_{ba}$. Using that parametrization the monopole field strength reads
\begin{equation}
	f= \theta_ae^a\wedge\tau - C_{ab}e^a\wedge e^b\,,
\end{equation}
which implies that under fracton gauge transformations $\delta \theta_a=\delta C_{ab} =0$. Generically, even in flat space those fields will be massive. Therefore, requiring $f=0$ we set them to zero. Finally, we end up we the set of gauge fields $(A_0,A_{ab})$, and using Eqs. \eqref{eq_transomegav} we conclude their transformation rule is 
\begin{equation}\label{eq_symmtrans}
\delta  A_{ab} = E^\mu\,_a E^\nu\,_b\nabla_\mu\nabla_\nu\lambda\,,\qquad \delta A_0 = -t^\mu\partial_\mu\lambda\,,
\end{equation}
with $\nabla_\mu\zeta_\nu = \partial_\mu\zeta_\nu - \Gamma^\alpha_{\mu\nu}\zeta_\alpha$, and the connection $\Gamma^\alpha_{\mu\nu} = t^\alpha\partial_\mu\tau_\nu + E^\alpha\,_a D_\mu e_\nu\,^a$. 

 The last necessary ingredient is an invariant volume form, that we defined as $vol_{n+1}\equiv\, ^\star 1=\sqrt{|G|}d^{n+1}x$.  With all this, a quadratic diffeomorphism and gauge invariant action for the theory can be expressed as\footnote{Notice that defining the volume as $vol_{n+1}=\sqrt{|G+B|}d^{n+1}x$ would produce higher order in fields corrections in the action. Nonetheless, we are interested in constructing a quadratic action.}
\begin{equation}\label{eq_gaugeAction}
    S = -\frac{1}{2}\int\,^\star (F_a +  z_b R^b\,_a)\wedge (F_a +  z_c R^c\,_a) + S_{SSB}[B_{\mu\nu}]\,.
\end{equation}

In fact notice, that if we take the flat space limit ($e^a= dx^a$, $\tau= dx^0$) the  action become  independent of $z_a$, which allows for massless fracton gauge fields, and we can safely write an invariant theory under the full fractonic gauge group
\begin{align}
    S &= -\frac{1}{2}\int\,^\star  F_a\wedge  F_a 
= \int \left[ F_{0 a b}F_{0 a b} - \frac{1}{2}F_{a b c}F_{a b c}\right]\,,
\end{align}
which has the form of a generalized electrodynamics theory, with the electric and magnetic fields being
\begin{align}
F_{0 a b} & = \partial_0 A_{ab} +\partial_a\partial_b A_0\,,\\
F_{a b c} & = \partial_a A_{bc} - \partial_b A_{ac}\,,
\end{align}
in full agreement with previous results \cite{Pretko:2016lgv}.

\section{Discussion}\label{sec:Discussion}

We have given a geometric interpretation to the symmetry group associated to the conservation of charge and dipole charge. In this picture the group is associated with a $2n+2$ dimensional space with the actual physical spacetime of dimension $n+1$ contained  in the larger space. The Nambu-Goldstone mode $\phi( x)$ appearing in the system can be understood as the breathing mode of the physical space inside the larger one. The advantage of that picture is that it allows us to construct consistently a gauge theory associated to the symmetry under discussion, in either flat or curved spacetime. Our results explain the incompatibility of the fractonic  symmetry with spatial curvature, since a fully invariant theory requires a Stueckelberg field. Therefore, we conclude that a fractonic system on a curved manifold will generically suffer spontaneous symmetry breaking due to curvature effects. This analysis pave the road for a more systematic analysis of theories preserving charges and their corresponding higher moments. In addition, it may help with the construction of low-energy effective fracton theories, because it provides a recipe to construct diffeomorphism and "gauge" invariant actions. Invariance will allow to derive covariant Ward identities for fracton charge, energy and momentum conservation. For instance, understanding how to couple the class of fractonic theories considered here to curved backgrounds, and the knowledge of the corresponding Ward identities could be fundamental to systematically construct fracton partition functions and in general hydrodynamics theories \cite{Banerjee:2012iz,Haehl_2015,Jensen,Glorioso}.

In the context of elasticity, this construction may help going beyond the current fractons/elasticity duality \cite{pretko2019crystal}. In fact, it would be interesting to explore within our geometric context the recently proposed generalization of such duality to the case of quasi-crystals \cite{Surowka:2021ved}. 

On the other hand, in quantum Hall systems, volume preserving diffeomorphisms have been related to the fractonic symmetry group discussed here \cite{du2021volume}. In fact, since the entire symmetry group preserve the two-form $dv$, it seems possible to connect our approach with volume preserving diffeos. However, an important difference between the two approaches is that the symmetric gauge field in \cite{du2021volume} are directly interpreted as a metric field, whereas, our fracton metric $B_{\mu\nu}$ depends quadratically on $ A_{ab}$.  Another interesting direction would be the construction of Chern-Simons actions. We leave the study of all these aspects for future investigations.

\emph{Acknowledgements} The author thanks Carlos Hoyos, Piotr Surowka, Kevin Grosvenor, and Karl Landsteiner for discussions. The author acknowledges financial support by the Deutsche Forschungs-gemeinschaft (DFG, German Research Foundation) under Germany’s Excellence Strategy through Würzburg-Dresden Cluster of Excellence on Complexity and Topology in Quantum Matter - ct.qmat (EXC 2147, Project Id 390858490), and the Norwegian Financial Mechanism 2014-2021 via the NCN POLS grant 2020/37/K/ST3/03390. 

\appendix

\section{Gauge Transformations}\label{App_GaugeTrans}
%%%%%%%%%%%%%%%%%%%%%%%%%%%%%%%%%%%%%%%%%%%%%%%%%%%%%%%%%%%%%%%%%%%%%%%%%%%%%%%%%%
%%%%%%%%%%%%%%%%%%%%%%%%%%%%%%%%%%%%%%%%%%%%%%%%%%%%%%%%%%%%%%%%%%%%%%%%%%%%%%%%%%

The gauge field associated to the fractonic symmetry group $\mathcal G$ discussed in the main text can be expanded as
\begin{equation}
   \tilde{ \mathcal A} =  \tilde\tau H + \tilde e^a P_a + \tilde \omega_a Q^a + s Q + \frac{1}{2}\omega^{ab}L_{ab}\,,
\end{equation}
and its corresponding curvature $\tilde{\mathcal R} = d\tilde{ \mathcal A} + \tilde{ \mathcal A}\wedge \tilde{ \mathcal A}$ reads

\begin{equation}
	\tilde{ \mathcal R} = d\tilde\tau H + D\tilde e^a P_a +  F_a Q^a  + ( d s + \tilde e^a\wedge\tilde \omega_a)Q + \frac{1}{2}R^{ab}L_{ab}\,,
\end{equation}
notice that the differential operator $D$ acts as a covariant exterior derivative, and is defined as $Dp^a  =  dp^a - \omega^a\,_b\wedge p^b$. In addition, the fractonic and rotational field strengths are
\begin{align}
	R^{ab} &=  D\omega^{ab}\,,\qquad	  F_a  = D\tilde \omega^a\, .
\end{align}

Since we are working with a non-Abelian symmetry group, under infinitesimal gauge transformations the gauge field and field strength transform  as  $\delta    \tilde{ \mathcal A}  = -    d\tilde{ \mathcal A}  - [   \tilde{ \mathcal A} ,\Lambda]$,  $\delta    \tilde{ \mathcal R}  =  - [   \tilde{ \mathcal R} ,\Lambda]$ respectively, with the gauge parameter
\begin{equation}
\Lambda = \zeta^0(x) H + \zeta^a(x) P_a +\beta_a(x) Q^a + \alpha(x) Q + \frac{1}{2}\beta^{ab}(x)L_{ab}\,.
\end{equation}

After some tedious but straightforward computation it is possible to derive the following set of transformation rules
\begin{align}
	\delta\tilde \tau &=  -d\zeta^0 \,,\\
	\delta \tilde e^a &= - D\zeta^a + \tilde e^b\beta_b\,^a \,,\\
	\delta \tilde \omega_a &= - D\beta_a + \tilde \omega_b\beta^b\,_a \,,\\
	\delta s =&  - d\alpha - \tilde e^a\beta_a + \tilde\omega_a\zeta^a\,,\\
	\delta \tilde \omega^{ab} &= - D\beta^{ab} \,,
	\end{align}
whereas the 	curvature fields transform as
\begin{align}
\delta(D\tilde e^a)  &=  R^a\,_b\zeta^b + D\tilde e^b\beta_b\,^a \,, \\
 \delta R^{ab} &=  D\omega^{ab}\,\\
\delta 	 F_a & = R_a\,^b\beta_b +  F_b\beta^b\,_a\,, \\
\delta (ds + \tilde e^a\wedge \tilde\omega_a) & =   -D\tilde e^a\beta_a + 	 F_a\zeta^a\,.
\end{align}

Notice that the transformation properties of the the fractonic gauge fields do not allow for a local gauge invariant action. Therefore, following \cite{isham1971nonlinear} we introduce Stueckelberg fields  to compensate such non-invariance. To do so, we introduce the Maurer-Cartan form, and its corresponding curvature
\begin{align}
\nonumber	\mathcal A &= \Omega^{-1}(d+\tilde{\mathcal{A}})\Omega =  \tau H +  e^a P_a + (\tilde\omega_a + D z_a) Q^a \\
 & +(\tilde v + z_a e^a + d\phi) Q + \frac{1}{2}\omega^{ab}L_{ab}\,,\\
\nonumber	 \mathcal R &=  d\tau H +  De^aP_a +  (F_a + z_b R^b\,_a)Q^a  +  (f + z_a De^a) Q \\
  &+ \frac{1}{2}R^{ab}L_{ab}\,.
\end{align}
where the new fields $\tau,e^a,\tilde v,f$ are defied as
\begin{align}
 	\tau & =  dx^0 +\tilde\tau \,,\\
 	 e^a &= Dx^a + \tilde e^a \,,\\
 	 s &=\tilde v + y^a\tilde\omega_a\,, \\
 	 f &= d\tilde v+  e^a\wedge \tilde\omega_a\,.
\end{align}

It is convenient to redefine the monopole gauge parameter as $\alpha =\lambda + y^a\beta_a$, such that gauge transformations act on the new fields as

\begin{align}
\delta\phi &=\lambda\,,\\
\delta  z_a  &=\beta_a + z_b\beta^b\,_a\,,\\
  \delta\omega^{ab} & = -D\beta^{ab} \,,\\
 \delta e^a &=  e^b\beta_b\,^{a}\,,\\
 \delta\tilde\omega_a &= -D\beta_a + \tilde\omega_b\beta^b\,_a \,,\\
 \delta F_a & = -\beta_a R^b\,_a + F_b\beta^b\,_a\,,\\ 
    \delta \tilde v  &= -d\lambda - e^a\beta_a\,,\\
 \delta  f  & =   -\beta_a De^a  \,.%\label{eq_transomegav}
\end{align}

\section{Constraints}\label{App_constraints}

For simplicity we set the  timelike and spatial torsions to zero
\begin{align}
	d\tau &= 0 \, \implies \tau = d(\mathrm{scalar}\, \,\mathrm{function})\,,\\
	De^a &=0\,\implies 	\omega_\mu\,^{ab} = \frac{1}{2}E^{\nu[a}\partial_{\mu}e_\nu\,^{b]} + \frac{1}{2}E^{\rho a}E^{\nu b} \partial_{[\nu}e_{\rho]}\,^c\,e_{\mu c}\,.
	\end{align}

On the other hand the inverse Higgs constraint \cite{low2002spontaneously,kharuk2018solving} together with the fixing $E^\mu\,_a \tilde v_\mu = 0$ imply
\begin{align}
	z_a & =  - E^\mu\,_a\partial_\mu\phi \,,\\
	\tilde v_\mu & = A_0\tau_\mu\,,
\end{align}
After such gauge fixing, the remaining gauge freedom is $\beta_a = -E^\mu\,_a\partial_\mu\lambda$.

The last constraint introduced is $f =0$, which fixes the dipole gauge field in terms of the monopole field $\tilde v$ modulo a symmetric tensor
\begin{equation}
    \tilde\omega_a = 2t^\mu E^\nu\,_a\partial_{[\mu}\tilde v_{\nu]}  \tau + \left(E^\mu\,_b E^\nu\,_a\partial_{[\mu}\tilde v_{\nu]}    +  A_{ab}\right)e^b\,,
\end{equation}
with $ A_{ab}= A_{ba}$.  Consistency of the gauge transformations of the dipole and monopole gauge fields demands
\begin{equation}
	\left[t^\mu\partial_\mu,E^\nu\,_a\partial_\nu\right] \lambda= 0\,\qquad \left[E^\mu\,_a\partial_\mu,E^\nu\,_b\partial_\nu\right]\lambda = 0\,,
\end{equation}
which is automatically satisfied due to the torsionless condition. Therefore, the symmetric field has the transformation rule
\begin{equation}\label{eq_symmtrans}
\delta  A_{ab} = E^\nu\,_aD_\nu(E^\mu\,_b\partial_\mu\lambda)= E^\mu\,_a E^\nu\,_b\nabla_\mu\nabla_\nu\lambda\,,
\end{equation}
with the diffeomorphism covariant derivative $\nabla_\mu\zeta_\nu = \partial_\mu\zeta_\nu - \Gamma^\alpha_{\mu\nu}\zeta_\alpha$, and the connection $\Gamma^\alpha_{\mu\nu} = t^\alpha\partial_\mu\tau_\nu + E^\alpha\,_a D_\mu e_\nu\,^a$. This covariant derivative satisfies
\begin{align}
  &  \nabla_\mu\tau_\nu  =0,\quad \nabla_\mu t^\nu =0\,,\\
  &  \nabla_\mu E^\nu\,_a + E^\nu\,_b\omega_\mu\,^b\,_a  =0\,,\\
  & \nabla_\mu e_\nu\,^a - \omega_\mu\,^a\,_b  e_\nu\,^b=0\,,
\end{align}
and the metric is covariantly constant $\nabla_\alpha G_{\mu\nu} =0$.

\bibliography{apssamp}% Produces the bibliography via BibTeX.

\end{document}